\documentclass{article}
\usepackage{amssymb}
\usepackage{amsfonts}
\usepackage{amsmath}
\usepackage{latexsym}

\usepackage{mathtools}
\usepackage{tikz, float}
\usepackage[british]{babel}

\usepackage{fancyhdr}
\usepackage[all]{xy}

\title{Marxism, logic,  and the law of the tendency for the rate of profit to fall}
\author{Robin Hirsch}

\begin{document}
\maketitle%\footnote{\nonumber Not for publication}
%\runningfooter{Not for publication}
 I am not on a campaign to demolish Marxism.  I am writing this note partly because I published a couple of articles \cite{hir:dialectics,H94m} relating to Marxism and, decades later, I now think I was wrong on a number of points, so I would like to clarify my own argument, mainly  for my own benefit.  But also, for political organisations aiming for the overthrow of Capitalism and particularly for such organisations where the importance of political theory is emphasised,  it seems to me that the adoption of a mistaken theory could be a serious handicap to their success.

One worry about Marxism, as Martin Hyland once pointed out to me, is that it is a theory of everything.  Marxism is a theory of history, politics, class-struggle, natural science, philosophy, religion and economics, but the underlying theory, \emph{dialectics}, claims to be a logic of change and interaction of processes in full generality.   The track record of theories of everything is not so good and while that in itself is not enough to rule Marxism out, it does place a greater burden of proof to provide a clear justification --- either empirical or logical -- for all its claims.  Because its claims are so ambitious,  we should require a higher level of proof than we do for, say, the licensing of a new drug.

\paragraph{Dialectics and Logic}
Marxists claim that classical logic is inadequate for reasoning about the complexities of the world and that \emph{dialectic} reasoning is required.  Dialectic reasoning is supposed to augment and generalise classical reasoning.    It is required in order to explain processes involving change and interaction, while formal reasoning is portrayed as only dealing with static events in isolation.  There are three laws of the dialectic: the law of unity, the negation of the negation, the transformation of quantity into quality, they derive from ancient Taoism, (but note that the three laws are not really laws, rather they are topic headings).  In materialist dialectics, based on the first two laws, contradictions exist not only in the realm of ideas, but in the real world (see \cite{Eng:83, Eng:77} for example, also \cite{Rees98}).  Accordingly, a clearly defined proposition can be both true and false at the same time.      Since contradictions in nature abound, any theory that does not contain contradictions must be partial, restricted, not capturing the essence. This contrasts with classical formal logic which adopts the law of non-contradiction.   Proof by contradiction,  where the deduction of a contradiction from a certain premise is used to prove that the premise is false (so not true), is based on the law of non-contradiction and is a  widely used  principle in natural science,  however it should be rejected by Marxists, since contradictions are held to be true.   Note that it's an elementary theorem of propositional logic that from one contradiction any proposition at all may be deduced. 

One advantage of this kind of dialectic  reasoning is that you can never be proved wrong.    If I assert some proposition $p$ and an opponent claims I'm wrong by giving a proof of not $p$, I can always reply, using dialectic logic, that I accept the proof of not $p$ but that $p$ is also true.    I consider this kind of reasoning to be profoundly illogical because it destroys the possibility of making reliable, irrefutable deductions.
Without reliable deductions, the  scientific method is lost.

\medskip

A working definition of  science is  evidence plus logic.  Accordingly, an argument is considered scientific if it is based on evidence and derives  its conclusions starting from the evidence, in logical steps.  There are different ideas about what counts as a logical step.  But in order for the argument to be scientific we must make explicit the justification for each logical step.

What counts as a logical argument may depend on the field.  But it should include a clear definition of any non-standard terms, any assumptions that are being made, and it should be clear how each statement or proposition follows from either evidence, from assumptions, or from what has gone before in the argument.  In formal logic, simple rules are given that allow you to deduce new statements from old (e.g. Socrates is a man, all men are mortal, \emph{therefore} Socrates is mortal).     In other branches of science a more informal approach is commonly accepted, however in a scientific theory each step in an argument must be clearly justified and it should be possible to formalise them if required.    If a step in the argument cannot be formalised  then it seems to me that the argument is unreliable and unscientific.   This is often  where theories go wrong.

In my view, a logical argument should not depend on domain specific knowledge, on intuition, nor on unspoken background knowledge.  In the syllogism about Socrates above, we do not need to know anything about Socrates, men or mortality.  However, a logical argument does require an understanding of very basic logical concepts, such as the meaning of `all', `some', `and', `equals', `not'.  In a logical argument we are not at liberty to interpret the word `not' as we choose, it  precisely excludes the possibility  that the argument that follows is true.  If we assert that $p$ and not $p$ is true, then at least part of what we say must be wrong.  For those who do not accept this view, that by this definition of `not' it is impossible for both $p$ and not $p$ to hold, a more anthropological analysis suggests that arguments based on contradictions are not convincing.  In  everyday conversation,  a contradiction in an argument is seen as a good reason to reject that argument (whereas, in Marxism contradictions are not only permitted, they are an \emph{essential part} of an argument\footnote{There are formal logics that deal with contradictions, e.g. \emph{paraconsistent logic}.  The intention here is to deal with inconsistent, contradictory information and try to resolve those inconsistency.  The idea is not that contradictions exist in reality, rather they consider databases of information which may not be reliably true.}).  An argument containing a contradiction is not persuasive.  When Matt Hancock argued that the UK was avoiding testing as a strategy for COVID in March 2020 and argued that this was because Science showed that testing would not save lives, and anyway the Government \emph{were} testing, the contradiction does not strengthen the argument, rather it undermines it.   Logically, the two parts of his argument cannot both be true to reality, hence something he said is clearly false.    

Of course, if we consider everyday conversational language, we also observe that contradiction frequently do appear.  If I ask ``is it hot in America?'' you may reply ``well, yes and no".  But the meaning I glean from your answer is not that there is a well-defined property that is true and false,  instead I conclude that my original question and its terms are unclear --- so that a simple yes or no answer would fail to describe the situation correctly.   What is being suggested is not a logical contradiction but a variation in the property, and a difficulty in defining what counts as hot\footnote{Quantum mechanics provides a more difficult challenge.  A  photon may `be' in two places simultaneously when unobserved, but after observation it may be in a single place.  However, there is no logical contradiction here, just a puzzle.}.   It seems to me that while apparent contradiction in an argument provide a useful way of identifying inconstencies in someone's beliefs and also may identify statements whose truth varies accoding to the situation or is not properly defined, but  a \emph{logical} contradiction in an argument always weakens it. 
 According to very simple laws of propositional logic, if we take both p and NOT p as assumptions we may draw any conclusions whatsoever.  Bertrand Russell may have said once  (although I cannot track down a reference) that the weaker the logic, the more surprising the conclusions.  

The theory  of dialectics fails to tell you when it is ok to conclude one thing from another, and when it is not.  Even so, there have been quite a few cases where deductions have been claimed, based on  the laws of dialectics   --- for example,  Hegel's proof that the number of planets could be only  seven, or  Marx's demonstration that the ellipse is a form of motion that realizes and resolves the contradiction between centrifugal and centripetal urges --- but these arguments cannot be scientific.   

 The underlying logic of Marxism, dialectics, is illogical and should be rejected.  At best it permits a kind of relativizism, at worst it prohibits logical reasoning.    Other aspects of Marxism retain value, however the foundation in an illogical system does pervade other aspects.

\paragraph{Value}
Marxism distinguishes itself from the wider political left by an interesting analysis of exploitation and profit.  The general anti-capitalist view is that capitalism is based on exploitation and is therefore sinful, being contrary to a natural concept of fairness.  Marx shares this view, but adds that exploitation is also the root cause of a problem for the Capitalist system that dooms it.   Exploitation leads to a rise in the organic composition of capital, which leads to a fall in the rate of profit.    This is the central content of the thesis elaborated by Marx  in the three volumes of Capital \cite{Marx:C}.  According to this thesis, the Capitalist system is a transitory system which cannot last long.  The reason that it cannot last long is that a mechanism built-in to the nature of Capitalism leads to a long term tendency for the rate of profit to fall.  This mechanism works as follows.    The rate of profit is defined as the ratio $\frac{S}{C+V}$, surplus value divided by total costs.  In order to improve productivity, over time the organic composition $\frac CV$ will rise and consequently (assuming some bound on the rate of exploitation $\frac SV$) the rate of profit will fall.

Here I will argue that the argument stated above is logical, given certain assumptions, but at least one of the assumptions (a bound on the rate of exploitation) is unjustified.    A consequence is that I do not know how long the Capitalist system will last.

\medskip

But first, lets examine some issues in this formulation of the theory.     In the statement ``The Capitalist system is transitory and cannot last long'' and in the phrase ``long-term tendency'', we might ask, what counts as long?  All previous economic systems have turned out to be transitory, so we might have guessed, without reference to the three volumes,  that Capitalism will end eventually, but Marx's thesis suggests more:  that  Capitalism will last significantly less time than previous systems, in particular significantly less than the thousand years of Feudalism.  After two centuries of Capitalism it can certainly be argued that insufficient time has elapsed for us to see this, on the other hand, the more time that passes without an end of Capitalism, the more problematic this is for Marx's thesis.

A second preliminary comment is that the theory makes use of quantities, like $C, V, S$, the rate of profit etc., and these quantities require an explanation in the theory --- an explanation that provides numerical values for these quantities.   Without such an explanation we cannot infer that a certain ratio of these quantities will fall.  
In Marx's writing there appear to be a number of types of value: exchange value, necessary labour time and  underlying value.  

  The \emph{exchange value} of a particular commodity tells you the ratio that you expect to exchange this commodity for other commodities, under average conditions.  Given certain assumptions about fluidity in the market, the exchange value of a commodity may be represented as a certain number in such a way that two commodities with the same number as their exchange value would represent a fair exchange, i.e. the sort of exchange you would expect to be able to make, in average conditions.   More generally, the ratio between exchange values of two commodities tells you (inversely) the ratio that you expect to exchange quantities of each commodity.    The monetary price of a commodity is an indication of its exchange value\footnote{Its an interesting problem to find conditions which are necessary, or sufficient, for a meaningful exchange value to be defined, and to construct other models which summarize the patterns of exchange of commodities in situations where a global exchange value cannot be defined.  I might look at that further some other time, here we may assume that an exchange value is well-defined.}.  On the face of it, the exchange value of a commodity should be observable and measurable, but as a result of enormous complexities relating to currency exchange rates, hidden costs and subsidies, it is very hard to establish  reliable exchange values in practice.

The \emph{necessary labour time} for a commodity is what it says, the duration of all the work needed to produce the commodity, measured in hours.  It has to be modified to take into account the fact that some types of labour are more valuable than others.  A skilled worker  may be able to produce more value in one hour than an unskilled worker, so an adjustment is needed to take this into account.  It gets complicated because the  skill is itself a commodity that may be purchased and the value of that skill would then be determined by the amount of labour needed for the necessary instruction of the worker.  There is a danger that we are already losing track of exactly what we mean by ``necessary labour time",  clarifying that would be an interesting project, and we will return to this later, however it is not our main concern for this note.

The \emph{underlying value} is the true value of a commodity, when distortions and illusions created by the system are removed.  It is not directly observable or measurable, but it is needed to understand the behaviour of exchange value, labour time and everything else.  It is the holy ghost of this value trinity\footnote{Unobservable, hidden parameters may well be required in order to explain how Capitalism works,  but any theory which adopts such parameters needs particular justification --- either empirical or logical --- of how these hidden parameters can be used to obtain a correct understanding of observable things, like the collapse of capitalism.  To illustrate a problem with the notion of a discrepancy between observed values and true underlying values, consider the well-known Newtonian equation $f=ma$.  Imagine a physicist postulated that the true law was $f=mv$ but because of certain peculiarities in the way we observe these quantities, we should multiply the right hand side of the equation by $\frac av$ to account for the illusion that is observable to us.   The true law is $f=mv$ but after adjustment it seems (superficially) that $f=ma$.  The argument would be discounted, if for no other reason that it falls foul of Occam's razor by introducing unnecessary and unobservable properties. }.  The \emph{Law of Value} is sometimes invoked to explain the connection between these different types of value, but the definition of this law of value varies considerably and is nowhere stated precisely.   
 
 One version of the Law of Value is the  \emph{Labour Theory of Value} (LTV).    According to this, the exchange value of a commodity is determined by the amount of labour that is required for its production.  
The problem, of course, is that LTV is empirically and systematically false.  Because of the equalisation of profit rates between different sectors of the economy, the exchange value of a commodity will differ from that predicted by LTV.  Marx was aware of this problem and referred to it as the \emph{transformation problem}\footnote{All the same, for much of volume 1 of Capital, Marx gives the strong impression that he is using the LTV.  In his laboriously worked examples he estimates that values produced by a worker over the ten hours of a working day, e.g. 4 hours for capital costs and depreciation costs, 3 hours for worker's own wage, 3 hours for the boss.}.
Marx argues that there is a transfer of underlying value from labour intensive industries to capital intensive industries, caused by the equalisation of profits.  This means that the exchange value of a commodity will differ from the necessary labour time.  
Since LTV is false, Marx's theory still requires an account of how  values are determined\footnote{It is true that Marx and marxists have written extensively about the relation between underlying value and exchange value.  Explanations are given as to why the rate of profit rose in  particular periods of history.  It might be that in a certain period, Capitalist were able to raise the rate of exploitation, or the organic composition did not rise, or perhaps that a chunk of value was stolen from one sector and deployed in another, somehow causing the rate of profit to rise.  Such arguments are based on hindsight.  What they do not give is a way of working out, given data about an economy,  how these quantities (exchange value, rate of exploitation, organic composition, etc.) will interact and how to work out what the rate of profit will be.}.

  One weak version of the law of value states merely that all value originates with labour.    The problem with such a qualitative weakening of the law is that Marx's thesis concerns numerically valued quantities, like the rate of profit, so the theory must explain and provide numerical values, and the weakened law of value says nothing about these.     Without a quantitative account of underlying value it is hard to see how it is possible to prove that the rate of profit will fall. 
  
  A stronger version of the law of value states that underlying value is determined by  necessary labour time, but makes no assertion about a relationship between exchange value and underlying value.
However, what counts for Capitalists when making investment decisions is the exchange value of the commodities they (have their workers) produce.  The true underlying value is of no interest to them, what matters is what they can exchange their commodity for.  Capitalists will happily invest if they expect that the rate of profit, measured by exchange value, is high, it won't trouble them one bit if the rate of profit, measured by underlying value, is low.  There is no reason why Capitalism should not continue happily in such conditions.

  Since it is exchange values that matter to capitalists, Marx's theory must explain why the rate of profit, measured as a ratio of actual exchange values, tends to fall if the organic composition rises.    Having adopted a definition of value let us state and examine the argument for the declining rate of profit.

Our assumptions are that (i) the organic composition $\frac CV$ rises without bound (ii) but the rate of exploitation is bound\,  say $\frac SV\leq b$, (iii)  $V>0$ and (iv) there is a positive lower bound on the rate of profit, below that bound Capitalism cannot continue, say $\frac {S}{C+V}>\epsilon$ for some $\epsilon>0$ so long as Capitalism continues.  Then
\[\frac{S}{C+V}\leq \frac{bV}{C+V}\leq b\frac {V}C\]
The first inequality follows from assumption (ii), the second follows from assumption (iii).  
Since $\frac CV$ is getting bigger without bound (assumption (i)) it follows that  eventually $\frac CV> \frac b\epsilon$.  At that point, $\frac S{C+V} \leq  b \frac{\epsilon}b<\epsilon$.  At this point, if not earlier,  Captalism must end, by assumption (iv).    It all works, its logical.  But note that if we drop  any one of our four assumptions, the conclusion does not follow. 

It may be of interest to note that the argument from the four assumptions to the conclusion that the rate of profit must fall to arbitrarily small levels, does not depend on the use of underlying value, nor of necessary labour time, it works equally well using exchange value as a measure.

The first assumption is that over time in order to raise productivity, investment will tend to shift into capital intensive forms of production, hence the organic composition of capital tends to rise. However,  productivity can also be increased by investing more intensely in the value of labour-power.  The value produced by a highly skilled worker in one hour can greatly exceed the value produced by an unskilled worker.   All investment in education and welfare can be considered as a cost of production counted as variable capital $V$.  Given that, it is not clear that there has to be a tendency for the organic composition of capital to rise.   Empirical evidence may help with this question, although even if historically the organic composition has risen, this would not prove that it will rise without any bound.

The second assumption, that there is an upper bound on the rate of exploitation, is also problematic.    If we are assuming that the organic composition of capital rises without bound it is easy to conceive that the rate of exploitation also rises without bound.  With a very high organic composition of capital a worker will be highly productive.  The cost $V$ of his/her labour need not be correspondingly high, it is only required that $V$ covers the cost of keeping the worker at a basic level of health, strength, skill and competence.  It is not clear why the rate of exploitation could not be arbitrarily high.    The third and fourth assumptions are not problematic.

The contrapositive of the argument above shows that assuming (i), (iii) and (iv) we may deduce that while Capitalism continues,  the  rate of exploitation will increase without bound.

  In summary, I find the various notions of value in Marxism confusing, in particular the so-called \emph{Law of Value} is most unclear.   The value that is well-defined (though hard or impossible to measure) and critical for the health of the economy is exchange-value.    Using exchange-value as a measure of value, Marx's argument for the declining rate of profit is a logical consequence of four assumptions.  However, two of those assumptions are doubtful and without all four of them we have no proof of a long term tendency for the rate of profit to become arbitrarily small.   

 I do not doubt that Capitalism, like all previous systems, will end.  It is clearly a dynamic system prone to booms and destructive slumps.    It is quite conceivable that a slump more severe than previous slumps will occur before very long, and that this would wreak enormous damage to our lives. 
  Whether Capitalism could recover from such a slump, perhaps in a new form, or whether a new economic system would emerge, remains open.  The nature of any new system that might arise from such a crisis is a matter of speculation.

\bibliographystyle{alpha}

 \def\www{/\allowbreak}

\end{document}